# Explanation of low efficiency droop in semipolar $(20\bar{2}1)$ InGaN/GaN LEDs through evaluation of carrier recombination coefficients

M. Monavarian,[1,a)] A. Rashidi,[1] A. Aragon,[1] S. H. Oh,[2] M. Nami,[1] S. P. DenBaars,[3] and D. Feezell[1]

[1] *Center for High Technology Materials, University of New Mexico, Albuquerque, NM 87106, USA.*
[2] *Department of Electrical and Computer Engineering, University of California, Santa Barbara, CA 93106, USA.*
[3] *Materials Department, University of California, Santa Barbara, CA 93106, USA.*

We report the carrier dynamics and recombination coefficients in single-quantum-well semipolar $(20\bar{2}1)$ InGaN/GaN light-emitting diodes emitting at 440 nm with 93% peak internal quantum efficiency. The differential carrier lifetime is analyzed for various injection current densities from 5 A/cm$^2$ to 10 kA/cm$^2$, and the corresponding carrier densities are obtained. The coupling of internal quantum efficiency and differential carrier lifetime *vs* injected carrier density (*n*) enables the separation of the radiative and nonradiative recombination lifetimes and the extraction of the Shockley-Read-Hall (SRH) nonradiative ($A$), radiative ($B$), and Auger ($C$) recombination coefficients and their *n*-dependency considering the saturation of the SRH recombination rate and phase-space filling. The results indicate a three to four-fold higher $A$ and a nearly two-fold higher $B_0$ for this semipolar orientation compared to that of *c*-plane reported using a similar approach [A. David and M. J. Grundmann, Appl. Phys. Lett. 96, 103504 (2010)]. In addition, the carrier density in semipolar $(20\bar{2}1)$ is found to be lower than the carrier density in *c*-plane for a given current density, which is important for suppressing efficiency droop. The semipolar LED also shows a two-fold lower $C_0$ compared to *c*-plane, which is consistent with the lower relative efficiency droop for the semipolar LED (57% *vs*. 69%). The lower carrier density, higher $B_0$ coefficient, and lower $C_0$ (Auger) coefficient are directly responsible for the high efficiency and low efficiency droop reported in semipolar $(20\bar{2}1)$ LEDs.

III-nitride light-emitting diodes (LEDs) are key elements in solid-state lighting and visible-light communication systems. The semipolar $(20\bar{2}1)$ orientation of GaN is of particular interest due to its substantially reduced polarization-related electric fields, narrow emission linewidth, robust temperature dependence, high indium incorporation efficiency, large optical polarization ratio, and low efficiency droop.[1–14] Recently, LEDs with very high output power (~ 1 Watt) and low efficiency droop were demonstrated using 12 to 14-nm-thick single-quantum-well (SQW) active regions on this orientation[2,15]. Theoretical investigations suggest that semipolar $(20\bar{2}1)$ LEDs have a ~3X larger electron-hole wavefunction overlap[3], which should predict larger $ABC$ recombination coefficients[16,17] and peak external quantum efficiencies that occur at higher current densities[18] compared to *c*-plane

---

a) Electronic mail: mmonavarian@unm.edu, morteza.monavarian@gmail.com,

LEDs. However, despite the promising properties and excellent device results for semipolar $(20\bar{2}\bar{1})$ LEDs, the reason behind the low efficiency droop is not well understood. The carrier dynamics have only been studied by time-resolved photoluminescence[19] and the *ABC* recombination coefficients have not been experimentally investigated at all. An investigation of the carrier dynamics and *ABC* recombination coefficients under electrical injection is useful to help explain the droop behavior, confirm previous theoretical investigations, and guide future LED design on semipolar $(20\bar{2}\bar{1})$.

In this work, we determine the differential carrier lifetime (DLT) and *ABC* recombination coefficients for semipolar $(20\bar{2}\bar{1})$ LEDs under electrical injection using methods similar to previously developed small-signal techniques[20–22]. Here we measure the small-signal frequency response to obtain the modulation bandwidth and extract the DLT. The DLT results are used in conjunction with internal-quantum efficiency (IQE) measurements to separate the radiative and nonradiative components of the lifetime. The recombination rates *vs* carrier density (*n*) are then computed in the context of an *ABC* model that includes phase-space filling effects. Fitting of the recombination rates *vs* carrier density then enables extraction of the *ABC* coefficients. The results indicate a three to four-fold higher *A*, a two-fold higher $B_0$, and a two-fold lower $C_0$ for this semipolar orientation compared to that of *c*-plane. To the best of our knowledge, this is the first systematic evaluation of the DLT and *ABC* parameters on a nonpolar or semipolar orientation using electrical injection.

The investigated sample is a semipolar $(20\bar{2}\bar{1})$ LED grown using metal-organic chemical vapor deposition (MOCVD) on a free-standing $(20\bar{2}\bar{1})$ GaN substrate from Mitsubishi Chemical Corporation. The active region of the LED consists of a single 12-nm-thick InGaN/GaN double heterostructure (DH) emitting around 440 nm. The LEDs consist of circular mesas with 60 μm diameter, indium-tin-oxide (ITO) p-contacts, and ground-signal-ground RF electrodes. Micro-LEDs are used to minimize the effects of the RC time constant and carrier transport in the evaluation of the DLTs.[23–26] Details of the epitaxial structure and device fabrication are described in Ref. 19 and Ref. 26.

The optical frequency response was measured using a microwave probe station and network analyzer. A small RF signal (100 mV) was combined with a DC bias using a bias tee. The light was collected from the top of the device into an optical fiber and focused on a silicon photodetector with a 2 GHz bandwidth. The received electrical signal was amplified using a 30 dB low-noise electrical amplifier. The amplified signal was then received by the network analyzer to determine the optical frequency response and 3-dB modulation bandwidth ($f_{3dB}$). The DLT was extracted from the 3-dB modulation bandwidth using $\tau_{\Delta n} = 1/(2\pi f_{3dB})$, where $\tau_{\Delta n}$ is the DLT. RC transport effects are neglected in this study based on their small contribution in micro-LEDs in previous studies[26]. However, in the future, RC transport effects can be included using more advanced rate equations and circuit modeling techniques[25]. Here we measure the optical frequency response under continuous-wave electrical injection since the temperature dependence of the optical frequency response and DLT are negligible based on the thermal study performed by David et al [27].

The shape of the IQE vs current density curve was determined by measuring the relative external-quantum efficiency (EQE) under pulsed (500 ns pulse-width, 10 kHz repetition rate, and 0.5% duty cycle) electrical injection to avoid self-heating effects. To adjust the relative IQE curve to an absolute IQE curve, we set the peak value to 93%, which was previously obtained for this sample using room-temperature/low-temperature photoluminescence (PL)[19]. We also assume the light-extraction efficiency is independent of current density.

Fig. 1 shows the DLT and IQE as a function of current density ranging from 5 A/cm² to 10 kA/cm². The DLT decreases with increasing current density and shows signs of saturation at high current densities (>8 kA/cm²) (Fig. 1(a)). The DLT ranges from 8.8 ns at 5 A/cm² to 0.42 ns at 10 kA/cm², which are lower than the reported values for the case of c-plane (~ 12 ns at 5 A/cm² to 0.5 ns at 10 kA/cm²)[21]. The IQE ($\eta_{IQE}$) increases with current density to a maximum of 93% at ~165 A/cm², after which the device experiences efficiency droop, reaching 40% at 10 kA/cm² (Fig. 1(b)).

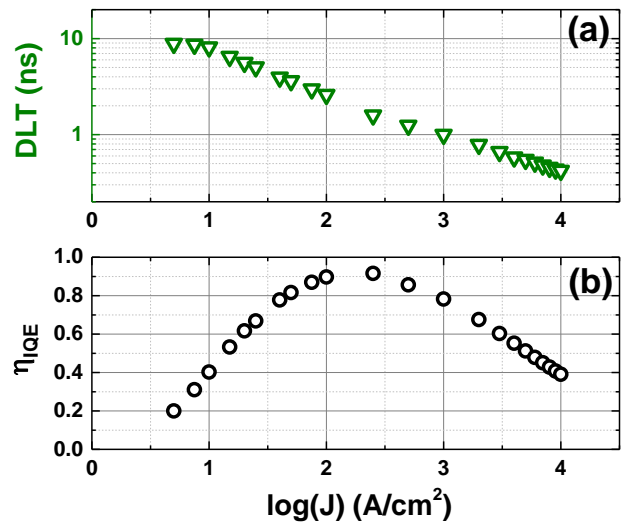

FIG. 1. (a) Differential carrier lifetime and (b) internal-quantum efficiency as a function of current density for the semipolar $(20\bar{2}1)$ LED.

Using the carrier lifetime as a function of current density, we then calculate the corresponding injected carrier density. The carrier density ($n$) is related to the DLT ($\tau_{\Delta n}$) and the generation rate ($G$) by the following [21,28]:

$$n = \int_0^G \tau_{\Delta n} \cdot dG. \tag{1}$$

For electrical injection, the generation rate $G$ is

$$G = \frac{J}{e \times d} \tag{2}$$

where $J$, $e$, and $d$ are the current density, electron charge, and active region thickness, respectively. The lower limit of the integral corresponds to a generation rate of zero, while the measureable minimum for the carrier lifetime corresponds to a non-zero generation rate. The carrier lifetime at zero generation rate is the Shockley-Read-Hall lifetime ($\tau_{SRH}$), which is typically obtained by extrapolating the lifetime at low $J$ to the y-axis. Here, $\tau_{SRH}$ is ~ 10 ns. Using (1) and (2), the carrier density corresponding to a given current density can be calculated with the knowledge of $\tau_{\Delta n}$. Fig. 2 shows this dependency for the semipolar ($20\bar{2}\bar{1}$) LED compared to the *c*-plane LED from Ref. 21. To reasonably compare the two orientations, we scale the carrier densities for the *c*-plane case up by 1.25 to account for the different active region thicknesses (12 nm semipolar *vs* 15 nm *c*-plane). Fig. 2 shows that the carrier density for a given current density is always lower in the semipolar case, which is due to the higher electron-hole wavefunction overlap in the semipolar case.[3] The difference in carrier density is particularly noticeable at low current densities before screening of the polarization field occurs. At higher current densities, the carrier densities become more similar.

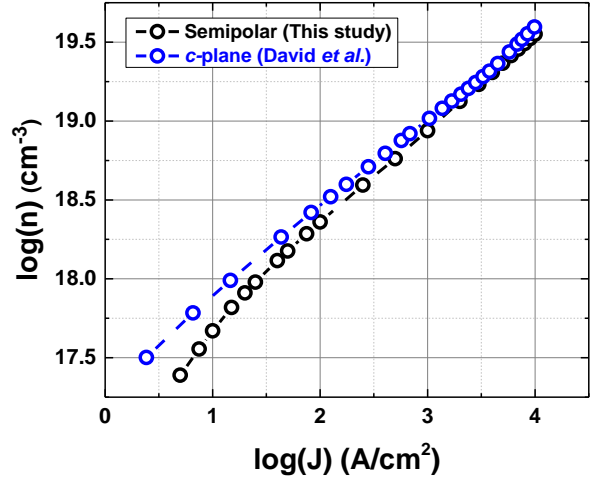

FIG. 2. (Color Online) Calculated carrier density ($n$) as a function of injected current density ($J$) for semipolar (black circles) and $c$-plane (blue circles) LEDs. The $c$-plane data is obtained from Ref. 21 and is adjusted to account for the difference in active layer thicknesses (12 nm semipolar $vs$ 15 nm c-plane).

Having $n$ as a function of $J$, we then plot $\eta_{IQE}$ and $\tau_{\Delta n}$ as a function of $n$ as shown in Fig. 3(a). The peak IQE occurs at a carrier density of $3\times 10^{18}$ $cm^{-3}$. Knowing $\tau_{\Delta n}$, $n$, and $\eta_{IQE}$, we find the radiative ($\tau_{\Delta nR}$) and nonradiative ($\tau_{\Delta nNR}$) recombination lifetimes separately using[21]

$$\tau_{\Delta nR}^{-1} = \frac{dG_R}{dn} = \eta_{IQE}\tau_{\Delta n}^{-1} + G \times \frac{d\eta_{IQE}}{dn} \qquad (3)$$

and

$$\tau_{\Delta nNR}^{-1} = \frac{dG_{NR}}{dn} = (1 - \eta_{IQE})\tau_{\Delta n}^{-1} - G \times \frac{d\eta_{IQE}}{dn}. \qquad (4)$$

The resulting radiative ($\tau_{\Delta nR}$), nonradiative ($\tau_{\Delta nNR}$), and total differential carrier lifetimes ($\tau_{\Delta n}$) as a function of carrier density are plotted in Fig. 3(b). At low $n$, $\tau_{\Delta nNR}$ approaches the SRH lifetime of $\tau_{SRH} \sim 10\ ns$ that was used to determine the lower limit of the integral in (1). Interestingly, the radiative ($\tau_{nR}$), nonradiative ($\tau_{nNR}$), and total PL lifetimes ($\tau_n$) obtained from previous optical measurements (see Ref. 19 for details of the measurement and calculations) follow very similar trends (Fig. 3(c)) to those obtained using the DLT approach (Fig. 3(b)). The DLTs are generally slightly lower than the PL lifetimes, as expected from the definition of DLT[29]. In addition, the optical excitation range was much lower for the PL case due to limitations in the setup, which results in an incomplete picture for the PL case (Fig. 3(c)).

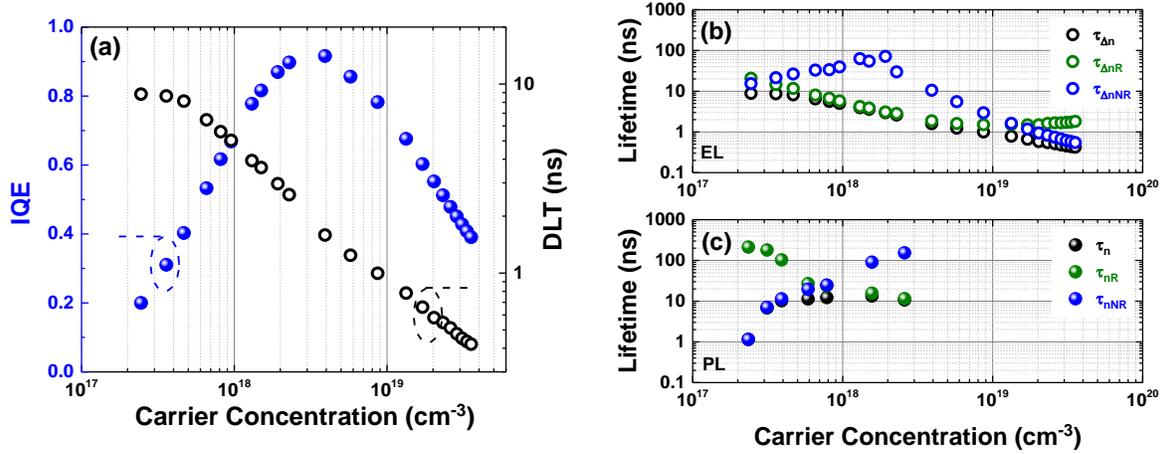

FIG. 3. (Color Online) (a) IQE and DLT vs. carrier density, (b) radiative, nonradiative, and total DLT vs. carrier density, and (c) radiative, nonradiative, and total PL lifetime vs carrier density.

Now following the work of David *et al.*[21] and considering an *ABC* model with a phase space-filling effects[20,30], the radiative ($G_R$) and nonradiative recombination rates ($G_{NR}$) are defined as:

$$G_R = G \cdot IQE = \frac{B_0}{\left(1+\left(\frac{n}{n^*}\right)\right)} n^2 \qquad (5)$$

$$G_{NR} = G \cdot (1 - IQE) = A(n)n + \frac{C_0}{\left(1+\left(\frac{n}{n^*}\right)\right)} n^3 \qquad (6)$$

where $A(n)$, $B_0$, and $C_0$ denote the Shockley-Read-Hall nonradiative, radiative, and Auger recombination coefficients, respectively. Also, $n^*$ is a fitting parameter accounting for the Fermi-Dirac carrier statistics in phase-space filling. To obtain the coefficients, we compute and plot three different parameters: $G_{NR}/n$ (Fig. 4(a)), $G_R/n^2$ (Fig. 4(b)), and $G_{NR}/n^3$ (Fig. 4(c)). The first parameter, at low carrier density, represents $A(n)$. The second parameter gives $B(n)$, from which $B_0$ and $n^*$ are obtained using fitting. The third parameter, at high carrier density, gives $C(n)$, from which $C_0$ is obtained knowing $n^*$ and $A(n)$. All of the recombination coefficients can be derived from the fittings shown in Fig 4. It is noteworthy that, unlike in the *c*-plane case[21], the *A* coefficient is a function of $n$, reaching a minimum at $n \sim 2 \times 10^{18} cm^{-3}$ (Fig. 4(a)). The value of $n$ at the minimum corresponds to the peak of $\tau_{\Delta nNR}$ in Fig. 3(b). This behavior is explained by the saturation of SRH recombination centers at high carrier density[19]. To obtain clean fittings to the data at low carrier densities, we use the original SRH model[31,32] described in Ref. 33 to describe the saturation. The SRH nonradiative lifetime is $\tau_{(SRH)} = \tau_{n0} + \tau_{p0}\left(\frac{n}{n+N_A}\right)$, where $\tau_{n0}$ and $\tau_{p0}$ are the electron and hole minority carrier lifetimes (which are defined as the inverse of the electron and hole capture coefficients[31]) and $N_A$ is the density of recombination centers. The choice of model for the saturation of SRH recombination is for fitting purposes only and does not affect the obtained values of the *ABC* coefficients since we extract (and compare with *c*-plane[21]) the *A* coefficient at the lowest carrier density (~ $3.5 \times 10^{17}$ $cm^{-3}$).

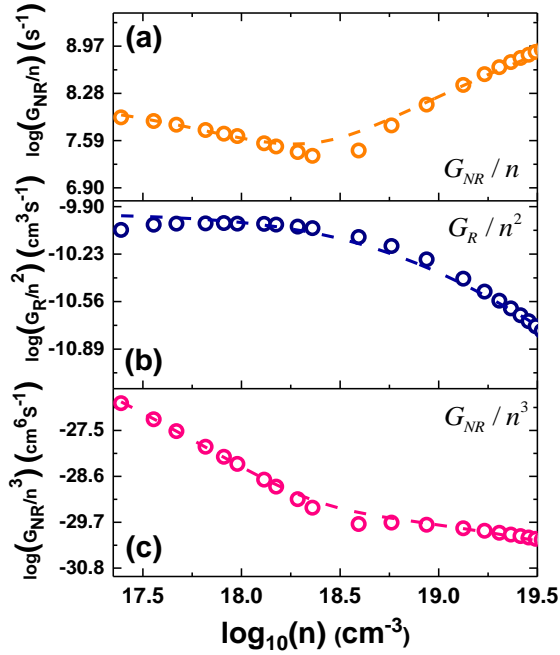

FIG. 4. (a) $G_{NR}/n$, (b) $G_R/n^2$, and (c) $G_{NR}/n^3$ as a function of carrier density calculated from the *n*-dependent radiative and nonradiative lifetimes. The dashed lines indicate the fitting by the *ABC* model.

A summary of the recombination coefficients, peak IQE, relative droop, and current density at the peak IQE is shown in Table I. For the sake of comparison, Table I also shows reference data reported using a similar technique for a *c*-plane LED with a single-quantum-well active region[21]. The semipolar LED shows some notable differences compared to the *c*-plane LED. First, the *A* and $B_0$ coefficients for the semipolar LED are higher by a factor of approximately four and two, respectively. The higher A and B coefficients are consistent with theoretical work by Kioupakis *et al*[16,17] predicting that the recombination coefficients are roughly proportional to the square of the electron-hole wavefunction overlap, which is about 2.5 to 3 times higher in semipolar ($20\bar{2}\bar{1}$) than *c*-plane for current densities below 500 A/cm$^2$.[3] The larger $A$ coefficient in the semipolar LED is not surprising based on the nature of the quantum well profile at low bias. The semipolar quantum well is nearly flat banded at low bias due to the cancellation of the built-in $p$-$n$ junction field and the polarization-related field.[3] The flat band profile leads to a large wavefunction overlap at low bias, which allows the first injected carriers to easily recombine through SRH centers at low carrier densities when the $A(n)n$ term is dominant. In $c$-plane, the first injected carriers at low bias have a poor wavefunction overlap, which should suppress SRH recombination and reduce the $A$ coefficient.

TABLE I. Recombination parameters for semipolar and polar InGaN/GaN LEDs

| Parameter | $(20\bar{2}\bar{1})$ (*This study*) | $(0001)$[a] |
|---|---|---|
| *Active Region* | Single | Single |
| *Well Thickness (nm)* | 12 | 15 |
| *Emission λ (nm)* | 440 | 430 |
| $\eta_{IQE}$[b] | 93% | 65% |
| $J_{peak}$[c] $(A/cm^2)$ | 165 | 50 |
| *Relative Droop* | 57% | 69% |
| $A$ $(s^{-1})$ | $7.6\times10^7$ | $2\times10^7$ |
| $B_0$ $(cm^3\ s^{-1})$ | $1.1\times10^{-10}$ | $7\times10^{-11}$ |
| $C_0$ $(cm^6\ s^{-1})$ | $4.3\times10^{-30}$ | $1\times10^{-29}$ |
| $n^*$ $(cm^{-3})$ | $6.3\times10^{18}$ | $5\times10^{18}$ |

[a] Data from Ref. 21
[b] $\eta_{IQE}$ is the peak IQE value.
[c] $J_{peak}$ is the current density corresponding to peak IQE.

Correcting for the difference in quantum well thicknesses between the two orientations, Fig. 2 shows that the semipolar orientation has a lower carrier density for a given current density, especially at low carrier densities where the internal electric field is not yet screened. This is a direct result of the larger $A$ and $B_0$ coefficients in the semipolar LED due to the reduced internal electric fields. A related effect was reported by David *et al.*[34] where the $A$ and $B$ coefficients were found to be significantly reduced in longer wavelength LEDs with larger internal electric fields. The lower carrier density in the semipolar LED is also responsible for increasing the current density corresponding to the peak IQE ($J_{peak}$). We observe a more than three-fold larger $J_{peak}$ in the semipolar LED despite the thinner active region (see Table I), which effectively delays the onset of droop effects to higher current densities.

In contrast to $A$ and $B_0$, the $C_0$ coefficient is lower in the semipolar LED compared to the $c$-plane LED (Table I). While the lower $C_0$ coefficient for the semipolar LED is not directly expected based on wavefunction overlap arguments alone, it is consistent with the lower relative efficiency droop in Table I and the higher value of $\frac{G_R/n^2}{G_{NR}/n^3}$ shown in Fig. 5, which does not result from an *ABC* model. Therefore, in addition to the wavefunction overlap, three additional factors that affect the Auger rate are briefly considered. First, at the high current densities relevant to Auger recombination, significant coulomb screening is expected in the active region and the difference in wavefunction overlap between semipolar and *c*-plane becomes smaller[35]. Thus, the difference in wavefunction overlap should have a smaller effect on the Auger rate than on the radiative rate. The convergence of the carrier density vs. current density curves at high current density in Fig. 2 supports this conclusion. Second, the Auger rate depends upon the detailed conduction and valence band structures, which constrain the allowed Auger transitions based on the requirements of energy and momentum conservation. Based on anisotropic strain, the band structures for semipolar and nonpolar differ significantly from those of *c*-plane[36–39], which may affect the Auger rate. On the other hand, indirect recombination based on alloy and phonon scattering was identified as the primary contributor to the Auger rate in InGaN[40] and was shown to be relatively insensitive to anisotropic strain[41]. Thus, the lower effective mass in the semipolar orientation should result in a higher $B$ coefficient but not significantly affect the $C$ coefficient if indirect Auger is the dominant process. Finally, localization as a result of indium alloy fluctuations has been shown to significantly affect the recombination coefficients, with $C$ being more strongly affected than $B$[42,43]. Some studies have shown low indium fluctuations in semipolar $(20\bar{2}\bar{1})$[6,44,45], which may result in the lower $C$ coefficient and higher $B$ coefficient observed in the semipolar case. Further experiments and first principles calculations specific to semipolar orientations are required to fully explain the origin of the lower Auger coefficient in semipolar $(20\bar{2}\bar{1})$.

The radiative and nonradiative rates can also be calculated without assuming an *ABC* model by using $G_R = G \cdot IQE$ and $G_{NR} = G \cdot (1 - IQE)$. Fig. 5 shows the ratio $\frac{G_R/n^2}{G_{NR}/n^3}$ as a function of carrier density. This ratio is essentially $B/C$ at high carrier densities and is approximately three times larger for the semipolar LED. The higher $\frac{G_R/n^2}{G_{NR}/n^3}$ in the semipolar case is important for realizing high-efficiency LEDs with low efficiency droop. Fig. 5 also shows the $B_0/C_0$ ratio calculated using the coefficients obtained from our fittings. As expected, the ratio $\frac{G_R/n^2}{G_{NR}/n^3}$ converges to $B_0/C_0$ at high carrier densities, further validating the fittings. The relatively constant value of $\frac{G_R/n^2}{G_{NR}/n^3}$ at high carrier densities for the *c*-plane case is consistent with a phase-space filling model where $n^*$ is the same for both $B(n)$ and $C(n)$. In the semipolar case, the slight negative slope in $\frac{G_R/n^2}{G_{NR}/n^3}$ at high carrier densities suggests the phase-space filling effects for $C(n)$ are slightly weaker than those for $B(n)$.

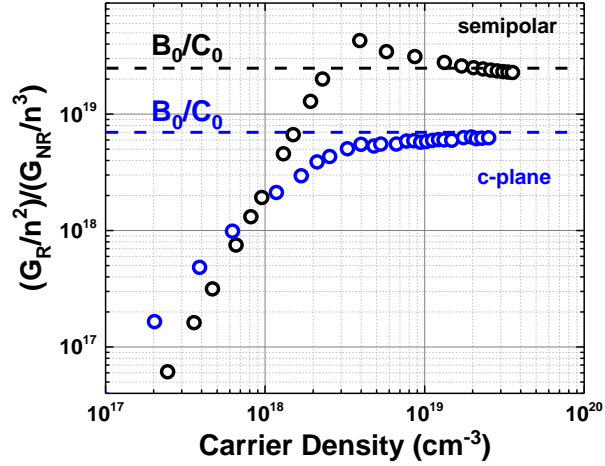

FIG. 5. (Color Online) The ratio $\frac{G_R/n^2}{G_{NR}/n^3}$ as a function of carrier density for semipolar (black circles) and $c$-plane (blue circles) from Ref. 21. The dashed lines indicate the extracted $\frac{B_0}{C_0}$ values from the fittings.

In summary, we report, for the first time, the recombination dynamics and *ABC* coefficients in electrically injected semipolar ($20\bar{2}\bar{1}$) InGaN/GaN single-quantum-well LEDs using a coupled differential carrier lifetime and internal-quantum-efficiency analysis. The radiative and nonradiative contributions to the differential lifetime as a function of carrier density were separated. In contrast to the nonradiative lifetime, the radiative lifetime initially decreases with $n$ until $1.3\times10^{19}$ cm$^{-3}$ and slowly increases afterwards. The behavior of radiative and nonradiative differential lifetime obtained by electrical injection agrees well with the corresponding lifetimes obtained from PL analysis. In the context of an *ABC* model with phase-space filling, the $A$ and $B_0$ coefficients are found to be higher by a factor of about four and two than those reported for a similar $c$-plane LED using the same technique. This is consistent with the lower carrier density for the same injected current density at low current density in the semipolar structure. The $C_0$ coefficient is lower by a factor of two in the semipolar LED, which is opposite to the first-order theoretical prediction based on wavefunction overlap arguments alone[16,17]. However, secondary effects such as coulomb screening at high current densities, band structure modifications, and the degree of carrier localization are also expected to affect the Auger coefficient and require further study in semipolar ($20\bar{2}\bar{1}$) quantum wells to fully explain the lower $C_0$ coefficient. The lower carrier density, higher $B_0$ coefficient, and lower $C_0$ (Auger) coefficient are directly responsible for the high efficiency and low efficiency droop reported in semipolar ($20\bar{2}\bar{1}$) LEDs.

This work was supported by Department of Defense award number W911NF-15-1-0428. The authors acknowledge the support from the Solid State Lighting and Energy Electronics Center (SSLEEC) at UCSB. The research carried out at UCSB made use of shared experimental facilities of the Materials Research Laboratory (MRL): an NSF MRSEC, supported by NSF DMR 1121053. The MRL is a member of the NSF-supported Materials Research Facilities Network (www.mrfn.org).

**REFERENCES**

[1] Y. Zhao, S. Tanaka, C.-C. Pan, K. Fujito, D. Feezell, J.S. Speck, S.P. DenBaars, and S. Nakamura, Appl. Phys. Express **4**, 082104 (2011).

[2] C.-C. Pan, S. Tanaka, F. Wu, Y. Zhao, J.S. Speck, S. Nakamura, S.P. DenBaars, and D. Feezell, Appl. Phys. Express **5**, 062103 (2012).

[3] D.F. Feezell, J.S. Speck, S.P. DenBaars, and S. Nakamura, J. Disp. Technol. **9**, 190 (2013).

[4] Y. Kawaguchi, C.-Y. Huang, Y.-R. Wu, Q. Yan, C.-C. Pan, Y. Zhao, S. Tanaka, K. Fujito, D. Feezell, C.G.V. de Walle, S.P. DenBaars, and S. Nakamaura, Appl. Phys. Lett. **100**, 231110 (2012).

[5] S. Marcinkevičius, R. Ivanov, Y. Zhao, S. Nakamaura, S.P. DenBaars, and J.S. Speck, Appl. Phys. Lett. **104**, 111113 (2014).

[6] S. Marcinkevičius, Y. Zhao, K.M. Kelchner, S. Nakamaura, S.P. DenBaars, and J.S. Speck, Appl. Phys. Lett. **103**, 131116 (2013).

[7] S.-H. Park and D. Ahn, J. Appl. Phys. **112**, 093106 (2012).

[8] Y. Zhao, S. Tanaka, Q. Yan, C.-Y. Huang, R.B. Chung, C.-C. Pan, K. Fujito, D. Feezell, C.G. Van de Walle, J.S. Speck, S.P. DenBaars, and S. Nakamura, Appl. Phys. Lett. **99**, 051109 (2011).

[9] C.-Y. Huang, M.T. Hardy, K. Fujito, D. Feezell, J.S. Speck, S.P. DenBaars, and S. Nakamaura, Appl. Phys. Lett. **99**, 241115 (2011).

[10] Y. Zhao, Q. Yan, C.-Y. Huang, S.-C. Huang, P.S. Hsu, S. Tanaka, C.-C. Pan, Y. Kawaguchi, K. Fujito, C.G. Van de Walle, J.S. Speck, S.P. DenBaars, S. Nakamaura, and D. Feezell, Appl. Phys. Lett. **100**, 201108 (2012).

[11] C.-C. Pan, T. Gilbert, N. Pfaff, S. Tanaka, Y. Zhao, D. Feezell, J.S. Speck, S. Nakamura, and S.P. DenBaars, Appl. Phys. Express **5**, 102103 (2012).

[12] Y. Zhao, S.H. Oh, F. Wu, Y. Kawaguchi, S. Tanaka, K. Fujito, J.S. Speck, S.P. DenBaars, and S. Nakamura, Appl. Phys. Express **6**, 062102 (2013).

[13] S. Marcinkevičius, R. Ivanov, Y. Zhao, S. Nakamura, and S.P. DenBaars, Appl. Phys. Lett. **104**, 111113 (2014).

[14] F. Wu, Y. Zhao, A.E. Romanov, S.P. DenBaars, S. Nakamura, and J.S. Speck, Appl. Phys. Lett. **104**, 151901 (2014).

[15] S.H. Oh, B.P. Yonkee, M. Cantore, R.M. Farrell, J.S. Speck, S. Nakamura, and S.P. DenBaars, Appl. Phys. Express **9**, 102102 (2016).
[16] E. Kioupakis, Q. Yan, D. Steiauf, and C.G.V. de Walle, New J. Phys. **15**, 125006 (2013).
[17] E. Kioupakis, Q. Yan, and C.G.V. de Walle, Appl. Phys. Lett. **101**, 231107 (2012).
[18] T. Melo, Y.-L. Hu, C. Weisbuch, M.C. Schmidt, A. David, B. Ellis, C. Poblenz, Y.-D. Lin, M.R. Krames, and J.W. Raring, Semicond. Sci. Technol. **27**, 024015 (2012).
[19] S. Okur, M. Nami, A.K. Rishinaramangalam, S.H. Oh, S.P. DenBaars, S. Liu, I. Brener, and D.F. Feezell, Opt. Express Vol 25 Issue 3 Pp 2178-2186 (2017).
[20] P.G. Eliseev, M. Osinski, H. Li, and I.V. Akimova, Appl. Phys. Lett. **75**, 3838 (1999).
[21] A. David and M.J. Grundmann, Appl. Phys. Lett. **96**, 103504 (2010).
[22] D.-P. Han, J.-I. Shim, and D.-S. Shin, Appl. Phys. Express **10**, 052101 (2017).
[23] S. Rajbhandari, J.J.D. McKendry, J. Herrnsdorf, H. Chun, G. Faulkner, Harald Haas, I.M. Watson, D. O'Brien, and M.D. Dawson, Semicond. Sci. Technol. **32**, 023001 (2017).
[24] R.X.G. Ferreira, E. Xie, J.J.D. McKendry, S. Rajbhandari, H. Chun, G. Faulkner, S. Watson, A.E. Kelly, E. Gu, R.V. Penty, I.H. White, D.C. O'Brien, and M.D. Dawson, IEEE Photonics Technol. Lett. **28**, 2023 (2016).
[25] A. Rashidi, M. Nami, M. Monavarian, A. Aragon, K. DaVico, F. Ayoub, S. Mishkat-Ul-Masabih, A.K. Rishinaramangalam, and D. Feezell, (submitted).
[26] A. Rashidi, M. Monavarian, A. Aragon, S. Okur, M. Nami, A. Rishinaramangalam, S. Mishkat-Ul-Masabih, and D. Feezell, IEEE Photonics Technol. Lett. **29**, 381 (2017).
[27] A. David, C.A. Hurni, N.G. Young, and M.D. Craven, Appl. Phys. Lett. **109**, 033504 (2016).
[28] D. Yevick and W. Streifer, Electron. Lett. **19**, 1012 (1983).
[29] L.A. Coldren, S.W. Corzine, and M.L. Mashanovitch, *Diode Lasers and Photonic Integrated Circuits* (John Wiley & Sons, 2012).
[30] D. Huang, J.-I. Chyi, and H. Morkoç, Phys. Rev. B **42**, 5147 (1990).
[31] W. Shockley and W.T. Read, Phys. Rev. **87**, 835 (1952).
[32] R.N. Hall, Phys. Rev. **87**, 387 (1952).
[33] A. Cuevas and D. Macdonald, Sol. Energy **76**, 255 (2004).
[34] A. David and M.J. Grundmann, Appl. Phys. Lett. **97**, 033501 (2010).
[35] E. Kuokstis, C.Q. Chen, M.E. Gaevski, W.H. Sun, J.W. Yang, G. Simin, M. Asif Khan, H.P. Maruska, D.W. Hill, M.C. Chou, J.J. Gallagher, and B. Chai, Appl. Phys. Lett. **81**, 4130 (2002).
[36] Y.-R. Wu, C.-Y. Huang, Y. Zhao, and J.S. Speck, in *Nitride Semicond. Light-Emit. Diodes LEDs* (Woodhead Publishing, 2014), pp. 250–275.
[37] H.-H. Huang and Y.-R. Wu, J. Appl. Phys. **107**, 053112 (2010).
[38] I.L. Koslow, M.T. Hardy, P. Shan Hsu, P.-Y. Dang, F. Wu, A. Romanov, Y.-R. Wu, E.C. Young, S. Nakamura, J.S. Speck, and S.P. DenBaars, Appl. Phys. Lett. **101**, 121106 (2012).
[39] S. Khatsevich and D.H. Rich, J. Phys. Condens. Matter **20**, 215223 (2008).
[40] E. Kioupakis, P. Rinke, K.T. Delaney, and C.G. Van de Walle, Appl. Phys. Lett. **98**, 161107 (2011).
[41] E. Kioupakis, D. Steiauf, P. Rinke, K.T. Delaney, and C.G. Van de Walle, Phys. Rev. B **92**, 035207 (2015).
[42] C. Jones, C.-H. Teng, Q. Yan, P.-C. Ku, and E. Kioupakis, ArXiv170206009 Cond-Mat (2017).
[43] T.-J. Yang, R. Shivaraman, J.S. Speck, and Y.-R. Wu, J. Appl. Phys. **116**, 113104 (2014).
[44] R. Shivaraman, Y. Kawaguchi, S. Tanaka, S.P. DenBaars, S. Nakamura, and J.S. Speck, Appl. Phys. Lett. **102**, 251104 (2013).

[45] Y. Zhao, F. Wu, T.-J. Yang, Y.-R. Wu, S. Nakamura, and J.S. Speck, Appl. Phys. Express **7**, 025503 (2014).